\documentclass{JHEP3}

\title{
On gravitational couplings in D-brane action }
\usepackage{epsfig}
\input{epsf}
\author{A. Fotopoulos$^{1,2}$ and A.A. Tseytlin$^{1,3,}$\thanks{Also at
Lebedev Physics Institute, Moscow.}
\\$^1$Theoretical Physics Group, Blackett Laboratory,
 \\Imperial College, London SW7 2BZ, U.K.
\\$^2$Centre de Physique Theorique,
\\Ecole Polytechnique, 91128 Palaiseau, France
\\$^3$ Smith Laboratory, The Ohio State University,
\\Columbus OH 43210, USA
}
\abstract{We compute
 the two closed string graviton -- two open string scalar
superstring scattering amplitude on the disc to
 show that there is no second-derivative curvature--scalar
coupling term  $R X^2$
in the low-energy effective action of a  D-brane  in curved space.
} \preprint{Imperial/TP/02 -3/3 \\
CPHT-RR 076.1102 \\
\hepth{0211101}} 
\begin{document}
%Declarations
\newcommand{\fig}[1]{Fig.~\ref{#1}}
\newcommand{\lp}{\left(}
\def \del {\partial}
\newcommand{\rp}{\right)}
\newcommand{\blp}{\biggl(}
\newcommand{\brp}{\biggr)}
\newcommand{\ze}{\zeta}
\def \ov {\over}

\def \ci{\cite}\def \foot{\footnote}
\def\be{\begin{equation}}
\def\ee{\end{equation}}
\def \k {\kappa}
\def \ov {\over}
\def \ha { { 1\ov 2}}
\def \bi {\bibitem}
\def \la{\label}
\newcommand{\rf}[1]{(\ref{#1})}
\def \m {\mu} \def\n {\nu} \def \s {\sigma} \def\t {\tau}
\def \a {\alpha} \def\b {\beta}
\newcommand{\apr}{\alpha'}
\newcommand{\al}{\alpha}
\newcommand{\bal}{\bar{\alpha}}
\newcommand{\si}{\sigma}
\newcommand{\de}{\delta}
\newcommand{\bpartial}{ \bar{\partial}}
\newcommand{\bz}{\bar{z}}
\newcommand{\bw}{\bar{w}}
\newcommand{\bh}{\bar{h}}
\newcommand{\bL}{\bar{L}}
\newcommand{\bp}{\bar{p}}
\newcommand{\bpsi}{\bar{\psi}}
\newcommand{\beps}{\epsilon}
\newcommand{\eps}{\epsilon}
\newcommand{\grad}{\bigtriangledown}
\newcommand{\cN}{{\cal N}}
\newcommand{\bx}{\bar{x}}

\def\appendix#1{
  \addtocounter{section}{1}
  \setcounter{equation}{0}
  \renewcommand{\thesection}{\Alph{section}}
  \section*{Appendix \thesection\protect\indent \parbox[t]{11.15cm}
  {#1} }
  \addcontentsline{toc}{section}{Appendix \thesection\ \ \ #1}
  }

%\start \vskip 2truemm
%%%%%%%%%%%%%%%%%%%%%%%%%%%%%%%%%
%Main text

%%%%%%%%%%%%%%%%%%%%%%%%%%%%%%%%%%%%%%%
\section{Introduction} \label{intro}
%%%%%%%%%%%%%%%%%%%%%%%%%%%%%%%%%%%%%%%

The structure of  gravitational coupling terms in
the   action  for a D-brane probe moving in a
curved space is of interest in many  contexts, in particular, in
brane-world constructions and gauge theory -- string theory duality.

The standard Born-Infeld \ci{Fradkin,Call}
action for a D-brane \ci{poll,POLL} in flat space \ci{LeighA}
has a direct  generalization to  curved ambient
 space\footnote{We absorb $2\pi \a'$ into the $U(1)$ gauge field strength.}
\begin{eqnarray}\label{DBI}
S_{DBI}= - T_{(p)} \int d^{p+1}\s \
 e^{- \phi(X)} \ \sqrt{ \det \big[ (G_{\mu
\nu}(X) + B_{\mu \nu}(X)) \partial_{\alpha} X^{\mu} \partial_{\beta}
X^{\nu} +  F_{\alpha \beta} \big] } + ...
\ .
\end{eqnarray}
Dots stand for various higher-derivative corrections
(present already in  flat space \ci{andr}).
We shall ignore the well-known WZ-type
 couplings to R-R potentials and concentrate on
 the parity-even part of the
action.

One may wonder still if the action \rf{DBI} correctly
describes the gravitational couplings of  the D-brane
even to the lowest order in derivatives.
A way to determine the precise form of
the low-energy D-brane effective action
is to compare the corresponding vertices with the
  open--closed string S-matrix
on the disc. Several of such studies checking the consistency of
\rf{DBI} and extending it to the next (4-th)  derivative order
were carried out in the past
\ci{malda,hashimoto,GarScatter,GarBI,BachasCurvature,Wyllard,foto}.
 It was found  (in agreement with expectation based on
 the leading
gravitational interaction term on the disc in type I string theory
being  $R^2$  \ci{TT}) that there is no Einstein-type $R$-term in
the D-brane action in type II superstring case\foot{$R$-term does
appear in the bosonic string D-brane action \ci{Corley,ardalan}.}
but there are  $O(\a'^2)$ correction  terms
\ci{BachasCurvature,foto,Wyllard}
 \begin{eqnarray} \label{LBachas}
&L^{(p)} =  T_{(p)}\ \sqrt {G}  \, e^{-\phi}  \bigg(
1-\frac{1}{24} \frac{ (4 \pi^2 \alpha' )^2  }{ 32 \pi ^{2}} \big[
(R_{T})_{\alpha \beta \gamma
\delta }(R_{T})^{\alpha \beta \gamma \delta } \nonumber \\
&-2(R_T)_{\alpha \beta }(R_T)^{\alpha \beta }-(R_{N})_{\alpha
\beta i j }(R_{N})^{\alpha \beta i j } +2 \bar{R}_{i j} \bar{R}^{i
j}\big]\bigg) \ .
\end{eqnarray}
Here $\a,\b,..=0,1,..., p$ are the ``parallel'' and
$i,j,..=1,...,9-p$ are the ``transverse'' indices and the tensors
$ R_{T}$ and $ R_{N}$ are constructed from the world-volume and
normal bundle connections and involve the second fundamental form
(see \cite{BachasCurvature,foto} for details). Note that
\rf{LBachas} cannot be written just in terms of the curvature of
the induced metric.

The analysis of \cite{BachasCurvature,foto} excluded  the standard
$R$ term  but it did not address the possible presence  of a
Brans-Dicke type  $R f(X)$ terms, e.g., $R X^2$, where $X$ stands
for the ``transverse'' ($X^i$) scalar components and $R$ is the
curvature in the directions parallel to the brane.\foot{One might
wonder whether the previous analysis of curvature corrections was
somehow incomplete. In \cite{foto} it was shown that the results
of \cite{BachasCurvature} had several ambiguities which had to be
fixed before one  could conclude that (\ref{LBachas}) is indeed
the correct action to order $\a'^2$. The case of the $RX^2$ term
has a similar ambiguity as we shall discuss below.}

One  may actually  rule out the presence of similar   terms
with explicit $X$-dependence on quite
general grounds.  For example,
 given a supergravity soliton, its effective action may be
derived in a static gauge. Since $X^i$ is a  goldstone boson
 reflecting the breaking  of
 translational invariance, the action should  depend on it only
through $\del X$.\footnote{There is also a functional dependence
of the background fields on $X=\bar X + \tilde  X $, e.g.,  $G_{\mu
\nu}(X) \simeq G_{\mu \nu}(\bar X )+
\tilde X^{i}\del_{i} G_{\mu \nu}(\bar X)+ \dots$,
but it  results only in terms with
 normal derivatives of  the background fields
which we are not interested in.}
Also, from the string-theory side,
the  $U(1)$ gauge invariance  combined with T-duality relation
between  $F_{\a \b}$ and $\del_{\a}
X_i$ implies that the effective action reconstructed from
string amplitudes should depend only on derivatives of
the embedding coordinates.  For example,  we may start
 with the
 D9-brane action which does not
contain transverse  $X$-scalars; its action is
  $\int d^{10} \s \ e^{-\phi} [ \sqrt{ \det ( G_{\a\b} +
F_{\a\b})} +\  $higher-derivative terms + $\ \a'^2 R^2$-terms $+ ...].$
Assuming that  the space-time metric is  flat in
some $9-p$ (toroidal) directions  we may
 apply T-duality in these directions. That will effectively
convert the corresponding components of $A_\a$ into $X_i$;
  we should then  finish with a Dp-brane action  in
a curved metric in ``parallel'' directions,
and  there is no way it can    contain $R f(X)$ terms.

%%%%%%%%%%%%%%%%%%%%%%%%%%%%%%%%%%%%%%%%%%%%%%%%%%%%%%%%%%%%%%%%

At this point one may wonder how the absence of the $RX^2$
term in D3-brane action is consistent with the AdS/CFT correspondence.
One  needs the standard conformal
coupling term  $\frac{1}{6} R X^2$ for the 6 scalars of the
$N=4$ SYM theory
to define the conformal stress tensor operator for the scalars,
so one would expect this term  to be present in a D3-brane probe action.
This apparent puzzle  was implicit already
 in \cite{GKT,GK,witten98,tseytlin98}.\footnote{One needs this curvature coupling  term to argue \ci{witten98}
that the moduli space is lifted when SYM theory is defined on a sphere.
This term is also  crucial  for conformally-invariant coupling
of SYM theory to external conformal supergravity sources \ci{tseytlin98}
needed to define the partition function form  \ci{GKP,Wi}
of the  AdS/CFT correspondence.}
%%%%%%%%%%%%%%%%%%%%%%%%%%%%%%%%%%%%%%%%%%%%%%%%%%%%%%%%%%%%%%%%%%%%%

This puzzle   was effectively resolved in \ci{sei}.
As was explained there, starting
with a {\it negative curvature    Einstein space  with  a conformal boundary}
 with
an arbitrary  curved boundary metric  $g_{\a\b}$
(e.g., asymptotically $AdS_{p+2}$ space),
 and considering a Dp-brane
probe placed close and {\it parallel}   to the boundary and
described by the {\it standard} DBI action \rf{DBI},
 one finds the effective conformal coupling term
 ${ p-2\ov 4 (p-1 )} R(g)  r^2$
for the normal-direction scalar $r$
($r^2 = X_i X_i$ in the $AdS_5 \times S^5$ case).
In more detail,
expanding the metric $g_{\a \b}$ in  geodesic distance
from the boundary\footnote{It is crucial here that eq. (3.6) in
\cite{sei} holds for the
 specific choices of the embedding and the
metric. Then  the radial coordinate $r$ can be identified with the
Riemann normal coordinate at the point we expand, and   the
expansion of the metric near  the boundary will have the form
$g_{\a \b}(x, r)= g_{\a \b}(x, 0) + \frac{1}{3} r^2 R_{r \a r \b}
+ \dots.$}
 and plugging the expansion  into the induced-metric determinant
part of the DBI  action for the
brane,  we get indeed  the $Rr^2$ coupling between
the boundary curvature and the normal scalar.
It should be stressed that this curvature coupling term  originates
not from an additional contribution to
 the DBI action but rather
from
 a specific choice of  (the expansion of) the   bulk metric,
 which is curved in both the  boundary
and the normal directions.
The  Weyl
invariance of the normal scalar
coupling to the curved boundary
%(i.e. ``world-volume'')
 metric is again  a consequence of the specific
embedding of the brane in the  AdS-type space with a  conformal
boundary.\footnote{In more general backgrounds or
for different orientations of the brane one  may end up with
 other effective couplings which may  not be Weyl-invariant.}

%%%%%%%%%%%%%%%%%%%%%%%%%%%%%%%%%%%%%%%%%%%%%%%%%%%%%%%%%%%%%%
%%%%%%%%%%%%%%%%%%%%%%%%%%%%%%%%%%%%%%%%%%%%%%%%%%%%%%%%%%%%%%%%%

To complement the above (already convincing) arguments,
we decided  to explicitly verify
 the absence of the  $RX^2$   term  in the
 D-brane action
by  a first-principle -- string S-matrix -- computation. The
previous discussions based on comparing the 3-point (one graviton
$h$  -- two scalar $X$) amplitudes are not sufficient for this
purpose. Here we are interested in the case when  the gravitons
are polarized along the D-brane world-volume
directions.\footnote{In general, in the case of the
``transversely'' polarized gravitons, there is an  ambiguity of
adding terms like $(D_i \Omega^{i}_{\a \b}) g^{\a \b} X^2$ and
$(D^i D_{\mu} \xi ^{\nu}_i) \Pi^{\mu}_{\ \nu} X^2$ where we have
used the normal frame $\xi^{\mu}_{i}$ and the second fundamental
form $\Omega^{i}_{\a \b}$\  ($\Pi^{\mu \nu}\equiv  \partial_\a
X^{\mu} \partial_\a X^{\nu}$).
The bulk covariant derivatives are
projected to the normal and tangent bundles using the projectors
$\de^{ij} \xi_i^{\mu} \xi_j^{\nu}$ and $\Pi^{\m\n}$
 \cite{BachasCurvature, foto}. Such
terms with appropriate coefficients can cancel the contribution of
$RX^2$ term to the $h XX$ amplitude without any influence on  the
four-scalar  amplitude.} In this case the $R X^2$ term gives
vanishing contribution to  $h XX$ amplitude.\footnote{Notice
also that the possible ambiguities mentioned in the previous
footnote give individually vanishing contributions and do not lead
to any contradiction.}

The direct way to rule out the $RX^2$ term is then
 to compute the
2-graviton -- 2-scalar  ($hhXX$) 4-point
superstring amplitude on the
disc
(with the Dp-brane boundary conditions)
and to compare it with the field-theory
amplitude  predicted by the sum of the  bulk Einstein
action and the DBI action \rf{DBI}.
This is what we are going to do below.
We shall start with the  superstring  computation in sections 2 and 3
and then show in  section 4 that the result
 is in complete
agreement  with the DBI action at
the second-derivative order.

%%%%%%%%%%%%%%%%%%%%%%%%%%%%%%%%%%%%%%%%%%%%%%%%%%%%%%%%%%%%%%%%%%
For  completeness, let us briefly mention some  other   work on
gravitational couplings on the brane. While there is no tree-level
$R$-term  on D-branes of type II superstring theories,  such term
may be induced at 1-loop string level in the case of reduced
amount of supersymmetry (see \ci{kiritsis01,Antoniadis02}).
 In the case of D-branes in  bosonic string
(and non-BPS branes in superstring) in addition to tree-level $R$-term
one  expects to find $R f(T)$ couplings
for the open string tachyon $T$
 \ci{Corley}.
In the case of branes in AdS case (in the  RS set-up \ci{RS})
one generically finds (by performing zero-mode analysis
\ci{garriga,chiba} which should be related to the
 discussion in \ci{sei})
Brans-Dicke type terms
on the 3-brane.

%%%%%%%%%%%%%%%%%%%%%%%%%%%%%%%%%%%%%%%%%%%%%%%%%%%%%%%%
\section{Preliminaries}\label{preliminaries}
%%%%%%%%%%%%%%%%%%%%%%%%%%%%%%%%%%%%%%%%%%%%%%%%%%%%%

As explained above,
we  intend to compute the tree-level  (disc) scattering
amplitude involving  two closed-string modes --  gravitons polarized
along the brane --
and  two open string modes --
 scalars $X$ describing transverse fluctuations of the brane.
This computation is aimed at
determining  whether a world-volume coupling $R X^2$ is present in
the D-brane action. Notice that the
contribution from such a term to the one graviton -- two scalar
amplitude
is vanishing as a result of the
lowest order equations of motion.

D-branes are non-perturbative string states \ci{POLL}. This means that they
are not part of the ordinary string spectrum and at   weak
coupling  have infinite mass compared to  string modes. In
space-time D-branes are represented as static p-dimensional
defects. As usual, that means that    we must impose different
(Neumann
and Dirichlet)
 boundary conditions on the   tangent and
 normal directions to the D-brane,
  \begin{eqnarray} \label{NeuDir}
  \partial_{\perp}X^{\alpha}\mid_{\partial \Sigma}=0 \nonumber \\
  X^{i}\mid_{\partial \Sigma}=0
  \end{eqnarray}
  The  Greek indices  ($ \alpha =0,1 \dots ,p $),
  correspond to the coordinates
 parallel to the brane (we shall call them ``world-volume'' directions)
 and the  Latin ones ($ i=p+1,
  \dots, 9$) -- to the  coordinates in the directions  normal
to the brane.

%When calculating tree string amplitudes we evaluate the partition
%function on a world-sheet with the topology of a disc.
Before
presenting details of our calculation, let us  review the basic
formalism of string vertex operators and their expectation values
on the disc. We follow closely the review of \cite{hashimoto} and
references there, in particular \cite{GarBI, malda}.
%\footnote{See also
%\cite{albertsson} for a recent approach to  taking into account
%supersymmetry.}

The string operators for an NS-NS
 massless closed string have the following general form
\begin{eqnarray}\label{GenOper}
V(z, \bar{z})= \epsilon_{\mu \nu} :V_{s}^{\mu}(z): \
:V_{s}^{\nu}(\bar{z}):
\end{eqnarray}
where $\mu=0,1,\dots,9$ and $s=0,-1$ denotes the superghost
charge or equivalently the picture in which the operator is defined.
The total superghost charge on the disk is required to be
$Q_{sg}=-2$ as a consequence of the
super-diffeomorphism invariance. The
holomorphic parts of the in the
pictures 0 and 1 are given  by:
\begin{eqnarray}\label{PicOper}
V_{-1}^{\mu}(p,z)= e^{-\phi(z)} \psi^{\mu}(z) e^{ i p \cdot
X(z)}\\
V_{0}^{\mu}(p,z)=( \partial X^{\mu}(z) + i p \cdot \psi(z)
\psi^{\mu}(z))e^{ i p \cdot X(z)} \nonumber
\end{eqnarray}
The Green's functions on the disc are found using the method of
image charges on a two dimensional surface \cite{Burgess}. Each
string vertex  inserted at position $z$ on the disc has an image at
$\frac{1}{\bz}$. Imposing Neumann or Dirichlet boundary conditions
we find the correlators on the disc
\cite{POL,hashimoto,klebanov,foto}:
\begin{eqnarray} \label{corrformulae}
&\langle \partial_z X^{\mu}(z) e^{i (Dp) \cdot X(\bw)} \rangle =
\frac{i (Dp)^{\mu} \bw }{1-z \bw} \nonumber \\
&\langle \partial_{\bz} X^{\mu}(\bz) e^{i p \cdot X(w)} \rangle =
\frac{i (Dp)^{\mu} w }{1-\bz w} \nonumber \\
&\langle \partial_z X^{\mu}(z) \partial_{\bw} X^{\nu}(\bw)
\rangle= \frac{ \eta^{\mu \nu}}{(1-z \bw)^2}\nonumber \\
&\langle \psi^{\mu}(z) \psi^{\nu}(w) \rangle = - \frac{\eta^{\mu
\nu}}{z-w} \\
& \langle \psi^{\mu}(z) \bar{\psi}^{\nu}(\bw) \rangle =  i
\frac{D^{\mu \nu}}{1-z \bw} \nonumber \\
&\langle c(z_1) c(z_2)c(z_3) \rangle = C_{D_2}^{ghost}(z_1-z_2)(z_2-z_3)(z_1-z_3) \nonumber \\
&\langle c(z_1) c(z_2) \bar{c}(\bz_3) \rangle= C_{D_2}^{ghost}
(z_1-z_2)(1-z_1 \bz_3)(1-z_2 \bz_3)\nonumber \\
&\langle e^{-\phi(z)} e^{-\bar\phi(\bw)} \rangle= \frac{1}{1-z
\bw}\nonumber
\end{eqnarray}
where $D_{\mu}^{\nu}$  reverses signs of the fields with Dirichlet
conditions and is defined in \cite{GarScatter} ($D^\lambda_\mu
D_{\lambda\nu} = \eta_{\mu \nu}$).

%AT
{ We shall
 specialize to
the case of a scattering involving only ``world-volume polarized''
gravitons. As we shall see in section \ref{FTA} the field theory
computation simplifies considerably for such a choice of
polarizations, mainly because in such a case the tadpole diagram
of a graviton with a ``transverse'' scalar is vanishing.}
We have
the following conditions on the momenta and polarizations due to
lowest order equations of motion and momentum conservation:
\begin{eqnarray}\label{kinematics}
&k_1+k_2+p+q=0 \ , \qquad{\rm Tr}(\eps_1)={\rm Tr}(\eps_2)=0\nonumber \\
&\eps_1 \cdot p=0\ ,   \qquad \eps_2 \cdot q=0  \\
& \ze_n \cdot p= \ze_n \cdot q=\ze_n \cdot k_n =0 \ .
\end{eqnarray}
Here $n=1,2$  labels  the two scalars with momenta $k^\mu_n$ and
polarizations $\zeta^\mu_n$ while graviton momenta are $p$ and $q$
and their traceless polarization tensors  have non-zero components
only along the world-volume directions of the brane. All momenta
are assumed to have only ``world-volume'' components  being
non-vanishing.

%%%%%%%%%%%%%%%%%%%%%%%%%%%%%%%%%%%%%%%%%%%%%%%%%%%%
\section{String theory amplitude}\label{stringampl}
%\paragraph{The correlator.}
%%%%%%%%%%%%%%%%%%%%%%%%%%%%%%%%%%%%%%%%%%%%%%%%%%%%%%%

%AT: I corrected  1/x factor  and -x into \bar x
%%%%%%%%%%%%%%%%%%%%%%%%%%%%%%%%%%%%%%%%%

The correlator we need to compute is:
\begin{eqnarray} \label{2cl+2o}
&A_{string} \propto  \int_{|z| \leq 1} d^2 \!z \, \int_{|x|=1} {d \!x \over
x}
\, \langle c(z') \bar{c}(\bz') V^{\mu \nu }_{(-1,-1)}(z',
\bar z') \, V^{\rho
\si}_{(0,0)}(z,\bar{z}) \nonumber \\
&(c(x)-c(\bar x))V^i_{(0)}(x)V^j_{(0)}(\bar x) \rangle \
 (\eps_1 D)_{\mu \nu} (\eps_2 D)_{\rho \si} \ze^1_i\ze^2_j\ .
\end{eqnarray}
Here we have fixed the $SL(2,R)$ Mobius symmetry gauge
by  setting  the position of one of the two gravitons to be at
the center of
the disc
$z'=\bar z'=0$ and the positions of the  two  open
string scalars to be at the  complex
conjugate points ($x, \bar x$) at the boundary of the disc.
 %on the circle }$|z|=1$
 \footnote{We  thank
Ashoke Sen for pointing out an error in $SL(2)$  gauge fixing in the original
 version of this paper.}
 % \footnote{ ??? The gauge fixing
%can be made more rigorous by changing the coordinates from
%$x_1,x_2$ for the scalars to $v_{\pm}=x_1 \pm x_2$ and then gauge
%fixing $v_+=0$.}

The correlators of the ghosts and superghosts are  (using
(\ref{corrformulae})):
\begin{eqnarray}
&\langle (c(x)-c(\bx)) c(0)\bar c(0)  \rangle = C_{D_2}^{ghost}(x-\bx) \nonumber \\
&\langle e^{-\phi(0)} e^{-\bar \phi(0)} \rangle= 1\nonumber
\end{eqnarray}
Since the scalars have polarizations only in the transverse
direction and all momenta are in world-volume directions
 the computation
simplifies considerably. From  $V^i_{(0)}V^j_{(0)}$  we need only
to retain the following correlators: \be\label{step1} \langle
\partial X^i \partial X^j \rangle = -\frac{\eta^{ij}}{(x-\bx)^2}  \ ,
\ \ \ \ \ \ \ \ \langle \psi^i \psi^j \rangle =
-\frac{\eta^{ij}}{(x-\bx)}\ , \ee with all other correlators
(coming from the cross product of world-sheet scalars and fermions
in  the $V^i_{(0)},V^j_{(0)}$ vertices)
 producing  vanishing contributions. It is easy to check that
given the symmetries of the $\eps_1,\eps_2$ tensors,  the part in
 $V^{\mu \nu }_{(-1,-1)}(z', \bar z') \,
V^{\rho\si}_{(0,0)}(z,\bar{z})$ multiplying  $\langle \partial X^i
\partial X^j\rangle $ in  (\ref{step1}) is vanishing. It remains then  to
compute the following correlator
\begin{eqnarray}
&\langle k_1 \cdot \psi(x) k_2 \cdot \psi(x) [ \psi^{\alpha}(0)
\bar{\psi}^{\beta}(0)\nonumber  \\
&(\partial X^{\gamma}\bpartial X^{\delta} + i\partial
X^{\gamma}q\cdot \bar{\psi} \bar{\psi}^{\delta} + i\bpartial
X^{\delta}q\cdot \psi \psi^{\gamma} - q\cdot \psi
\psi^{\gamma}q\cdot \bar{\psi}
\bar{\psi}^{\delta})] \nonumber \\
&[ e^{2i k_1 X}e^{2i k_2 X}e^{i p X}e^{ip \bar{X}}e^{i q X}e^{i q
\bar{X}}]\rangle \label{step2}
\end{eqnarray}
which should  be multiplied with $\langle \psi^i \psi^j \rangle$
in (\ref{step1}) and $(\eps_1 D)_{\mu \nu} \to (\eps_1)_{\alpha
\beta}$, $(\eps_2 D)_{\rho \si} \to (\eps_2)_{\gamma \delta}$.

After some tedious computations and using the symmetries of the
polarization tensors and the symmetry under interchange of the
two scalars  we
arrive at  the following result
for the integrand in \rf{2cl+2o}
(we ignore overall numerical coefficient and
%AT
isolate the
polarization tensor factors
$\epsilon_1,\epsilon_2, \zeta^1, \zeta^2$):
%AT
\begin{eqnarray}\label{step3}
{\cal A}_{string} &\sim  i   |1- \bx z|^{4q\cdot k1}|1- x \bz|^{4q\cdot k2}
|z|^{2p\cdot q} (x-\bx)^{4 k_1 \cdot k_2 +1} \eta ^{ij} \\
&\bigg[ k_1^{\alpha}k_2^{\beta} (k_1^{\gamma}k_1^{\delta} \ I_1
+ k_1^{\gamma}k_2^{\delta} \ I_2 ) \nonumber \\
&- k_1^{\alpha}k_1^{\gamma}(k_2 q \ \eta^{\beta
\delta}-k_2^{\delta}
q^{\beta}) \ I_3 \nonumber \\
&- \frac{k_1^{\alpha}k_2^{\gamma}}{2}(k_2 q \ \eta^{\beta
\delta}-k_2^{\delta} q^{\beta}) \ I_4 -
\frac{k_1^{\gamma}k_2^{\alpha}}{2}(k_1
 q \ \eta^{\beta \delta}-k_1^{\delta} q^{\beta}) \ I_5 \nonumber \\
&+ \frac{1}{2}(k_1  q \ \eta^{\alpha \gamma}-k_1^{\gamma}
q^{\alpha})(k_2 q \ \eta^{\beta \delta}-k_2^{\delta} q^{\beta})\
I_6 \bigg] + (1\leftrightarrow 2)\nonumber
\end{eqnarray}
Here  $(1\leftrightarrow 2)$ stands for  symmetrization under
interchange of the two scalars and the two graviton polarizations
and momenta,  and \be \label{integrands} I_1= \frac{|1+\bx
z|^2}{|1-\bx z|^2|z|^2} \ , \qquad I_2=\frac{(1- |z|^2)^2+ |z|^2
(x^2+ \bx ^2) - (z^2+ \bz^2)}{|1-\bx z|^2|1-x z|^2|z|^2}, \ee $$
I_3= \frac{(1+\bx z)(1-x \bz)^2(1-xz) \bx - \ c.c.}{(x-\bx)|1-\bx
z|^2|1-x z|^2|z|^2} \ , \qquad I_4= \frac{(1-x z)(1+x \bz) \bx - \
c.c.}{(x-\bx)|1-\bx z|^2|z|^2}  $$
$$ I_5= \frac{(1-x \bz)(1+x z) \bx - \
c.c.}{(x-\bx)|1-x z|^2|z|^2}\ ,  \qquad I_6= \frac{((1-|z|^2)^2}{|1-x
z|^2|1-\bx z|^2|z|^2} \ . $$
%In \rf{step3} we have  not included
%overall numerical coefficient.
% and the  factors of string coupling constant
%which appear in the definition of the vertex operators.

%\section{Extracting the poles}\label{poles}
%\paragraph{Extracting the poles.}
Next, we should perform the  integration over the
world-sheet coordinates $x$ and $z,\bar z $. Since we are
interested only in determining  the $R X^2$ contribution, we do
not actually need to compute the whole integral: it is  sufficient
to  extract the terms with only two powers of  the momenta.
 Given that all the  terms in \rf{step3} have four powers of
momenta
% it is easily understood that
we should just extract the residues of the poles on the
disc.\footnote{There is a  possibility that we have residues of
poles which do not diverge as we take the limit $\alpha ' \to 0$.
These lead to higher order contact terms and can be ignored.} The
integrand has various poles. The integral over $z$  has poles at
$z=0,x,\bx$. There are no poles when $x \to \bx$ as we can easily
infer from (\ref{step3}),(\ref{integrands}) except possibly  when
$z \to x \to \bx \to  \pm 1$. This latter  case will be discussed
later in this section.

We need to keep in mind now that when we are integrating
poles at $x, \bx$, residues should be taken with factor 1/2 since
they are at the boundary of the integration region. As a first
step we expand each of the integrands  in (\ref{integrands})
 in  some small region around each pole of $z$.
The residue is extracted using the standard rule (see, e.g.,
\cite{bern}, eq. (4.7)):
%{\it This is a triviality, I would not make ref to Bern here}
\begin{eqnarray}
 \int dy \ y^{-1+\apr k_i k_j} \to -\frac{1}{\apr k_i k_j} \ .
% \qquad (\apr \to 0)
\end{eqnarray}

 For example,  let us consider $I_1$ in \rf{integrands} which has poles at
$z=0,x= {1 \ov \bar x }$. Expanding around a given pole both $I_1$ and the terms in
the first line of  (\ref{step3}) we get:
\begin{eqnarray}\label{example}
\int {dx \over x} (x-\bx)^{4k_1 \cdot k_2 +1}\int_{0}^{2\pi}
d\theta  \int_{r> \epsilon} dr \ r \
 r^{2pq-2} \to \frac{8 \pi}{2pq}\ ,
\end{eqnarray} $$ \int {dx \over x}(x-\bx)^{4k_1 \cdot k_2+4 q \cdot k_2 +1}
\int_{0}^{\pi} d\theta  \int_{r> \epsilon} dr \  r \
 |1+1|^2 r^{4q \cdot k_1-2} \to \frac{8 \pi}{2q \cdot k_1}\ . $$

We have introduced an UV cut-off $\epsilon$ which we take to zero
when we extract the poles. We take the limit $\apr \to 0$, i.e.
expand in powers of momenta,  for all exponents which do not
contain singular terms.\footnote{Keeping these contributions would
produce  only  higher order corrections.} We have also accounted
for the relative factor $\frac{1}{2}$ between poles at the
boundary and in the interior by integrating $\theta$ only from $[
0 , \pi]$ in the second of (\ref{example}). Notice that the
$x=e^{i \varphi}$ integral has decoupled in the pole regions from
the one over $z$ and  is equal to $\int_{0}^{\pi} d \varphi \sin
\varphi= 2$ for all the cases in (\ref{example}).\footnote{The
integration region is chosen to be $[0, \pi]$ since the remaining
region up to $2\pi$ corresponds to exchange of the open string
states which was taken into account explicitly in (\ref{step3}).}

%The  reader might object
One may worry  that there also poles at  $z \to
\bx$ or $x \to \bx$ in the expression above. These potential
singularities could appear  at the points of the disc where the
three operators approach each other, $z=x= \bar x = \pm 1$.
%AT
As we   argue in Appendix A
  there are no massless poles in these cases other than
those considered in (\ref{example}).
%{\bf here goes the Appendix}

%\paragraph{The String amplitude.}
Following the same procedure  for all terms in
the amplitude and symmetrizing with respect to the two scalars we
get:
\begin{eqnarray}\label{stramp}
& A_{string} \sim (k_1 \eps_1 k_2) [ -(k_1 \eps_2 k_1)
\frac{qk_2}{4(qk_1)(pq)}-(k_2 \eps_2 k_2) \frac{qk_1}{4(qk_2)(pq)}
+ (k_1 \eps_2 k_2) \frac{1}{2pq} ] \nonumber\\
&-\frac{1}{4pq}[(k_1 \eps_1 \eps_2 k_1)(k_2q)+(k_2 \eps_1 \eps_2
k_2)(k_1q)-(k_1 \eps_1 q)(k_1 \eps_2 k_2)-(k_2 \eps_1 q)(k_1
\eps_2 k_2)]  \\
&+ \frac{k_1q}{4(k_2q)(pq)}[(k_1 \eps_1 \eps_2 k_2)(k_2q)-(k_1
\eps_1 q)(k_2 \eps_2 k_2)] + \frac{k_2q}{4(k_1q)(pq)}[(k_2 \eps_1
\eps_2 k_1)(k_1q)-(k_2 \eps_1 q)(k_1 \eps_2 k_1)]\nonumber \\
&+\frac{1}{4pq}[{\rm Tr}(\eps_1 \eps_2) (k_1q)(k_2q) - (q \eps_1 \eps_2
k_2)(k_1q)- (q \eps_1 \eps_2 k_1)(k_2q)+(q\eps_1 q)(k_1 \eps_2
k_2)]+ (1 \leftrightarrow 2) \nonumber
\end{eqnarray}
where $(1 \leftrightarrow 2)$
stands for the remaining
symmetrization in graviton polarization and momenta.
The apparent ``double-pole'' momentum factors
($1 \ov (qk_1)(pq)$, etc.)
here can be eliminated
by using momentum conservation.

As a check, one may demonstrate that  the  result is
gauge invariant under $(\eps_i)_{\mu \nu} \to (\xi_i)_{\mu} p_{\nu} +
(\xi_i)_{\nu} p_{\mu}$
(the first and fourth  lines of (\ref{stramp}) each are separately
 gauge invariant, while  the
second and third lines combine into a gauge-invariant expression).

Using conservation of momentum to eliminate some $q$-momenta in
terms of $k_1$ and $k_2$, and  symmetrizing in the graviton
polarizations,  we arrive at a much simpler  expression:
\begin{eqnarray}\label{stramp2}
&A_{string} \sim 2(\ze_1 \ze_2) \bigg[ (k_1 \eps_1 k_2)(k_1 \eps_2
k_2)\frac{1}{2pq} + (k_1 \eps_1 \eps_2 k_2)\frac{k_1q}{2(pq)}+
(k_2 \eps_1 \eps_2 k_1)\frac{k_2q}{2(pq)}
\nonumber \\
&+(k_1 \eps_1 k_1)(k_2 \eps_2 k_2)\frac{k_1q}{4(k_2q)(pq)} + (k_2
\eps_1 k_2)(k_1 \eps_2 k_1)\frac{k_2q}{4(k_1q)(pq)}+{\rm Tr}(\eps_1
\eps_2) \frac{(k_2q)(k_1q)}{4(pq)} \bigg]
\end{eqnarray}

%%%%%%%%%%%%%%%%%%%%%%%%%%%%%%%%%%%%%%%%%%%%%%%%%%%%%
\section{D-brane field theory amplitude}\label{FTA}
%\paragraph{The field theory amplitude}.\\
%%%%%%%%%%%%%%%%%%%%%%%%%%%%%%%%%%%%%%%%%%%%%%%%%%%%%%%

Let us now compare the string-theory result  with
field theory amplitude that follows from
%particular
 graviton--scalar interaction terms
in D-brane action.

%We shall
%follow the discussion  (and conventions) in  \cite{GarBI}.

The full field-theory  action contains the standard bulk
supergravity action (written in the Einstein frame) \be S_{10} = -
{ 1 \ov 2 \kappa^2} \int d^{10} x  \ \sqrt { -g } \ [ R - \ha (
\del \phi)^2  + ... ] \ , \la{ggg} \ee and the D-brane action. The
low energy dynamics of  a Dp-brane  in a curved space
 is  encoded in the DBI action.
The D-brane action  contains  the world-volume
 massless scalar fields $
X^{\mu} (\sigma)$ (the embedding coordinates
of the p-brane in the ambient
space-time)
and the $U(1)$ gauge fields
$ A_{\alpha}(\sigma)$
coupled to the bulk supergravity fields
($\sigma^{\alpha}$ are  the world-volume
coordinates).
 In addition,  there are supersymmetric partners of these
fields but they will be irrelevant for our present discussion.
We shall also ignore the WZ type  term describing
well-known   coupling of D-brane to
 the R-R potentials.
In terms of the Einstein-frame bulk metric $g_{\m\n}$
the DBI action may be written as
\begin{eqnarray} \label{dbi}
S_{DBI}= -T_{p} \int d^{p+1} \! \sigma \,  \ e^{\frac{p-3}{4}
\phi } \sqrt{ -\det[\tilde{g}_{\alpha \beta}(X)  + e^{-{\phi/2}}
\tilde{B}_{\alpha \beta}(X) +
e^{-\phi/2}F_{\alpha \beta}]} \ ,
\end{eqnarray}
where
$ \tilde{g}_{\alpha \beta}= g_{\mu\nu}(X) \del_\alpha  X^\mu \del_\beta X^\nu
 $ , $ \tilde{B}_{\alpha \beta} = B_{\mu\nu} (X)
\del_\alpha  X^\mu \del_\beta X^\nu
$
are the pull-backs on the brane of the corresponding bulk tensors.

Our aim is to compute the 2-graviton ($h$)-- 2-scalar ($X$) tree amplitude
and to compare it with string-theory result, to see
if we need to introduce some additional second-derivative
interactions to
the DBI action  like $R X^2$.

The field-theory amplitude will contain contact $hhXX$
contribution from the DBI action,  a scalar
exchange between the  two $hXX$ vertices,
 and also   a graviton exchange
between the closed string  vertex $hhh$ present in the
 bulk supergravity action
 and the  $hXX$ vertex present
in the DBI action (see Fig.1). Since there is no bulk coupling for
the dilaton of the form $\phi hh$ there will be no dilaton
exchange diagram, i.e. the dilaton (and $B_{\m\n}$) contributions
may be ignored in the present case. The scalar--gauge field
couplings are also irrelevant, i.e. we may set  $A_\a=0$. {An
even more significant simplification is that our choice of momenta
and polarizations for the graviton excludes mixing vertices like
$\partial^\alpha X^i h_{\alpha i}$. These "tadpole" diagrams would
have complicated considerably the field theory computation since
we would have to include diagrams with two and even three
intermediate states.}

 Expanding the induced metric
in \rf{dbi} near the flat space and using the static gauge $X^\a =
\sigma^\a$ we get ($\a=0,...,p; \ i=1,...,9-p$)
\begin{eqnarray} \label{GPB1}
\tilde{g}_{\alpha \beta}= g_{\alpha \beta}(\s,X)  + 2 g_{i(\alpha} (\s, X)
\partial_{\beta)} X^i + g_{ij}(\s,X)
\partial_{\alpha}X^i
\partial_{\beta}X^j
\end{eqnarray}
 \begin{eqnarray} \label{Fluc}
g_{\a \b}= \eta_{\a \b} + 2 \kappa h_{\a \b}(\sigma,X)\ , \ \ \ \ \
g_{ij}= \delta_{ij} + 2 \kappa h_{ij}(\sigma,X)\ , \ \ \ \ \
g_{\a i}=  2 \kappa h_{\a i }(\sigma,X)\ . \
\end{eqnarray}
In what follows we shall assume that the graviton is polarized
parallel to the brane, i.e. $h_{ij}=0=h_{\a i}$.

%%%%%%%%%%%%%%%%%%%%%%%%%%%%%%%%%%%%%%%%%%%%%%%%%%%%%%%
%\begin{figure}
\FIGURE{
%\begin{center}
%\epsfxsize=10cm \epsfbox{FT.eps}
\epsfig{file=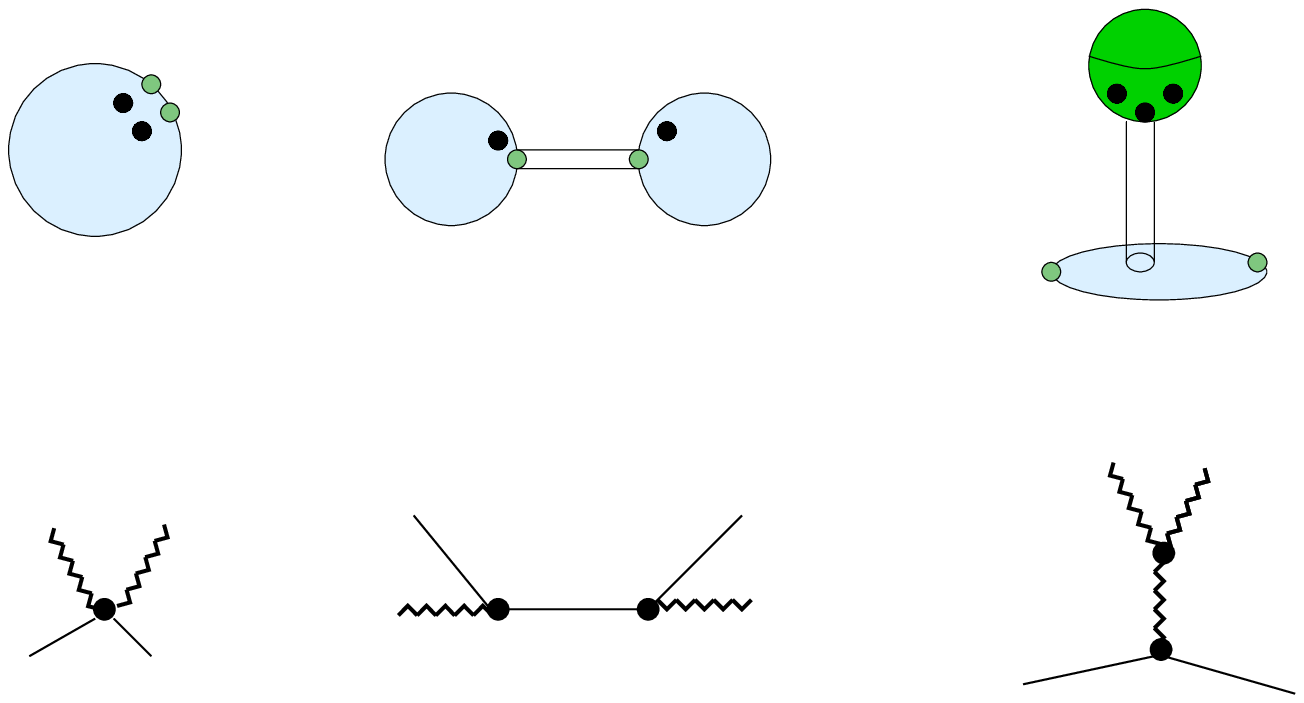}
%\end{center}
\caption{This  figure shows  three types
of contributions to the scattering amplitude,
both in string-theory (upper raw)
 and field-theory (lower raw) representation.
In string diagrams darker  dots stand for the  graviton
vertices  and
lighter dots for the scalar vertices; the double line is an
open or closed string propagator. In field-theory
diagrams the wiggly  lines are  graviton and straight lines are
scalar propagators.
 The
first contribution  is a contact term (coming from the
 region when all 4 points on the disc are close to each other).
 The second one is the s-channel exchange
contribution originating from the factorization of the disc into
two discs connected by  a scalar propagator (corresponding
to the region where points come  close  pairwise).
 The third diagram corresponds
to the region where the points of graviton insertions are close to
each other so that the amplitude factorizes to a sphere and a disc
connected by a graviton propagator. }}\label{fig:FT}
%\end{figure}
%%%%%%%%%%%%%%%%%%%%%%%%%%%%%%%%%%%%%%%%%%%%%%

To expand the square root of the determinant in \rf{dbi} we use
formula
\begin{eqnarray}
&\sqrt{\det(\delta_{\; \beta}^{\alpha}+ M_{\; \beta}^{\alpha})}= 1
+ \frac{1}{2}
 M_{\; \alpha}^{\alpha}-\frac{1}{4}  M_{\; \beta}^{\alpha} M^{\beta}_{\; \alpha}
 +\frac{1}{8}  (M_{\; \alpha}^{\alpha})^2 \nonumber \\
 &+ \frac{1}{6}M_{\; \beta}^{\alpha} M^{\beta}_{\; \gamma}M^{\gamma}_{\; \alpha}
 - \frac{1}{8}M_{\; \beta}^{\alpha} M^{\beta}_{\; \alpha}M^{\gamma}_{\; \gamma}+
 \frac{1}{48}(M^{\alpha}_{\; \alpha})^3\dots
\end{eqnarray}
%The vertices are found by multiplying by  $i$ the corresponding
%terms in the Lagrangians.
Using Fourier representation for  the fluctuations
$h_{\alpha \beta}$ and $X^i$
and expanding \rf{dbi}
we  get the following scalar propagator
and $hXX$ and $hhXX$  vertices (multiplying them by $i$)
\begin{eqnarray}
&P_{i} ^{j} = -\frac{ i \delta _{i}^{j}}{k^{2}} \\
&V^{h}_{XX}= -2i \kappa T_{p} (\ze_1 \ze_2)(k_1 \eps k_2) \\
&V^{hh}_{XX}= 4i \kappa^2 T_{p} (\ze_1 \ze_2)\big[  (k_1 \eps_1
\eps_2 k_2) +
(k_2 \eps_1 \eps_2 k_1)-{1\ov 2} (k_1k_2){\rm Tr}(\eps_1\eps_2) \big]\label{DBIvertex} \ .
\end{eqnarray}
To compute the graviton exchange contribution we will need also
the $hXX$ vertex with off-shell graviton (which we shall denote by
$H_{\m\n}$, with polarization tensor $E_{\m\n}$) \be V^{H}_{XX}=
-2i \kappa T_{p} \bigg[ (\ze_1 \ze_2)(k_1 E k_2) (\ze_1 \ze_2) -
{1 \ov 2}  (\ze_1 \ze_2)(k_1k_2)E^{\mu}_{\mu} -  (k_1k_2) (\ze_1 E
\ze_2)\bigg]  \ . \la{xxh} \ee The graviton propagator
corresponding to the bulk Einstein action is
\begin{eqnarray}\label{GrPro} ({P}_H)_{\mu \nu, \lambda \rho}=
-\frac{i}{2p^2 }(\eta_{\mu \lambda}\eta_{\nu \rho}+\eta_{\mu
\rho}\eta_{\nu \lambda}-\frac{1}{4} \eta_{\mu \nu}\eta_{\lambda
\rho})\ ,
\end{eqnarray}
while the vertex for the two on-shell gravitons and one off-shell
one is \cite{GNW}:
\begin{eqnarray}\label{3GrV}
&V(H,h_1,h_2)=-4 \kappa (H_{\mu \nu}-\frac{1}{2} \eta_{\mu \nu}
H^{\lambda}_{\lambda})
[h_{\rho \si} R^{(1)}_{\rho \mu \si \nu} \   \nonumber \\
&-\frac{1}{4} h_{\rho \si,\mu}h_{\rho \si,\nu}+\frac{1}{2}h_{\rho
\mu,\si}(h_{\si \nu,\rho}-h_{\rho \nu,\si})] \ ,  \\
&R^{(1)}_{\rho \mu \si \nu}=\frac{1}{2}(h_{\nu \rho,\mu
\si}+h_{\mu \si,\nu \rho}-h_{\mu \nu,\si \rho}-h_{\rho \si,\nu
\mu})\ . \nonumber
\end{eqnarray}
The relevant field-theory amplitude contributions are shown in Fig. 1.
%\fig{fig:FT}.
The contact contribution is given by the
$V^{hh}_{XX}$ term in (\ref{DBIvertex}):
\be
(A_c)^{hh}_{XX}= i \kappa^2 T_{p} (\ze_1 \ze_2)
\bigg[ 4 (k_1 \eps_1
\eps_2 k_2) + 4
(k_2 \eps_1 \eps_2 k_1)-2(k_1k_2){\rm Tr}(\eps_1\eps_2) \bigg] \ . \la{one}
\ee
The contribution from the  exchange of a scalar field (s-channel)
is:
\be
(A_s)^{hh}_{XX}=(V^i)^{h}_{XX}P^j_i(V_j)^{h}_{XX}=2 i \kappa^2
T_p\bigg[ (k_1 \eps_1 k_1)(k_2 \eps_2 k_2)\frac{1}{k_2q}+(k_1 \eps_2
k_1)(k_2 \eps_1 k_2)\frac{1}{k_1q}\bigg] \ ,  \la{two}
\ee
where we have symmetrized in both scalars and gravitons.
Finally,
it is long but straightforward to compute the contribution from
the exchange of a graviton (t-channel)
using \rf{xxh},(\ref{GrPro}) and (\ref{3GrV}):\footnote{It turns out that
the
last term in $V^H_{XX}$ does not actually
contribute  due to the
structure of the three-graviton vertex.}
\begin{eqnarray}
&(A_t)^{hh}_{XX}=(V^{\mu \nu})^{H}_{XX}(P_H)_{\mu \nu,
\lambda \rho}V^{\lambda \rho}(H,h_1,h_2)=i \kappa^2 T_p\bigg[-2(k_1
\eps_1 \eps_2 k_2)(\frac{k_1q}{pq}+1)\nonumber \\
&-2(k_2 \eps_1 \eps_2
k_1)(\frac{k_2q}{pq}+1)-{\rm Tr} (\eps_1\eps_2)\frac{(k_1q)(k_2q)}{pq}-(k_1\eps_1k_1)
(k_2\eps_2k_2)\frac{k_1q}{(k_2q)(pq)}
\nonumber \\
&-(k_1\eps_2k_1)(k_2\eps_1k_2)\frac{k_2q}{(k_1q)(pq)}
-\frac{1}{pq}(k_1\eps_2k_2)((k_1\eps_1k_1)+2(k_1\eps_1k_2)+(k_2\eps_1k_2))\nonumber\\
&+\frac{1}{pq}(k_1\eps_1k_2)((k_2\eps_2k_2)+(k_1\eps_2k_1))
+(\eps_1 \leftrightarrow \eps_2)\bigg]\ . \la{thr}
\end{eqnarray}
Combining \rf{one},\rf{two} and \rf{thr}
and using the symmetry under
$\eps_1 \leftrightarrow \eps_2$ in $(A_t)^{hh}_{XX}$ to simplify
some terms,  it is possible  to show that the field theory
amplitude reproduces the string-theory
 amplitude (\ref{stramp2}).
This rules out
 the presence of an extra $RX^2$ term in the DBI action.

The only  potential caveat could be the following.  Since we
 have not been careful to include  the normalization factors
in the string amplitude,  the  agreement
is  only up to the overall coefficient.
One could imagine
 that   $RX^2$ term could produce a contribution
which is also  proportional to the full string amplitude (\ref{stramp2}).
However,
 this possibility is excluded  as  $RX^2$ cannot give
an s-channel contribution present in
the string amplitude.\footnote{It may   still contribute to the
t-channel since it  modifies  the $V^H_{XX}$ vertex.}
One may also try to add
some other terms which could account for an additional
s-channel contribution; the only candidate with the right
number of momenta is $X_i \Omega^{i \alpha}_{\alpha}$,
where $\Omega^{i}_{\alpha \beta}$ is the second fundamental
form \cite{BachasCurvature,foto}. Such term, however,
 is  proportional to
the lowest-order  equations of motion and therefore can be removed
by a
field redefinition.

We conclude that the standard DBI action \rf{dbi}
is in full agreement with the string S-matrix at the second
 derivative order.

%Acknowledgments
\acknowledgments{We  would like to thank S. Kachru for raising
 the question  about possible  presence
  of $RX^2$ term in the D-brane action
which motivated the present work. We are also grateful to I.
Klebanov for extensive discussions and collaboration at an initial
stage of this project. We also thank P.Vanhove for a useful
advice. We are grateful to A. Sen for pointing out an error in an
earlier version of this paper. A.A.T. acknowledges the hospitality
of the Aspen Center for Physics while this work was in progress.
 The work of
A.A.T.  was supported in part
 by the grants DOE DE-FG02-91ER40690,   PPARC SPG 00613,
INTAS  99-1590,  and by  the Royal Society Wolfson Research Merit
Award. The work of A.F. was supported in part by the European
Commission under contracts HRPN-CT-2000-00131 and
HRPN-CT-2000-00148.

%%%%%%%%%%%%%%%%%%%%%%%%%%%%%%%%%%%%%%%%%%%%
\setcounter{section}{0}

%%%%%%%%%%%%%%%%%%%%%%%%%%%%%%%%%%%%%%%%%%%%%%%
\appendix{On singularities of the string
amplitude}\label{appendix}
%%%%%%%%%%%%%%%%%%%%%%%%%%%%%%%%%%%%%%%%%%%%%%%%%%

Our aim  here  will be to show
that the contribution to the amplitude from the
region  in the integration
 space where the three vertex operators (of graviton at point $z$
 and two scalars at points $x$ and $\bar x$)
  approach each other at the same time  is finite. We
will  study the integral from the $I_1$ term in
(\ref{integrands}): \begin{equation}\int {dx \over x} \int d^2 z
|1- \bx z|^{4q\cdot k1}|1- x \bz|^{4q\cdot k2} |z|^{2p\cdot q}
(x-\bx)^{4 k_1 \cdot k_2 +1}\frac{|1+\bx z|^2}{|1-\bx z|^2|z|^2}
\end{equation}
where $x=e^{i \varphi}$.
 We want to examine the integral for $z \to x$. Let us expand the
 integrand around $z=x+ \rho e^{i \theta}$ where $\rho$ is radial
 distance from the boundary of the disc and $\theta \in [ 0 ,
 \pi ]$. We want the leading contribution for $\apr \to 0$,
 i.e. to lowest order in momenta,
  so we
  take the momenta  to zero for any term in the
integrand as long as this does not produce a singularity. We get this
way:
\begin{equation} \label{expansion}
-4\int_0^\pi d \varphi \int_0^\pi d \theta \int_{\rho> \epsilon}
d\rho\ \rho^{4 q \cdot k_1-1} (\sin \varphi )^{4 k_1 \cdot k_2 +1}
|1-e^{2 i \varphi}+ \rho e^{i \varphi + \theta}|^{4 q \cdot k_2}
\end{equation}
where $\epsilon$ is a radial cut-off with the upper limit of the
radial integration not specified but assumed to be small enough for our
approximation to make sense. Now we can split the integration over
$\varphi$ into two regions. One far from $\varphi=0$ and another
close to it. The latter corresponds to the limit where $x \to \bx$
which we wish to examine and we will cut-off the singular region
by the same parameter
 $\epsilon$ as for the radial
direction. It is   more rigorous to  split the integration as:
$\int_\epsilon^{\pi \over 2} = \int_{\epsilon}^A +
\int_A^{{\pi\over 2}}$, where $A \ll {\pi \over 2}$ is an
irrelevant constant which should drop out  at the end.
  We also
integrate only up to $\varphi= {\pi \over 2}$ since there is also
the antidiametric point $\varphi=\pi$ which we need to treat in a
similar manner. The evaluation in the first region gives:
\begin{equation} \label{region1} -4\int_A^{{\pi \over 2}} d
\varphi\int_0^{\pi}d \theta \int_{\rho> \epsilon} d\rho \rho^{4 q
\cdot k_1-1} (\sin \varphi )^{4 k_1 \cdot k_2 +1} \to 8\pi\cos A
({1 \over 4 q \cdot k_1} + \log \epsilon) \sim 8 \pi  ({1 \over 4
q \cdot k_1} + \log \epsilon)\end{equation} where in the last
expression we expanded for small $A$ and kept only the
$A$-independent terms. For the second region we can expand the
integrand in (\ref{expansion}) for small $\varphi$ as well:
\begin{equation}\label{region2}\int_\epsilon^A d
\varphi\int_0^{\pi} d \theta \int_{\rho> \epsilon} d\rho \rho^{4 q
\cdot k_1-1}  \varphi ^{4 k_1 \cdot k_2 +1}|\rho^2+4\varphi^2-4
\varphi \rho \sin \theta|^{4 q \cdot k_2} \end{equation} Now we
can change coordinates from ``cartesian" $(\rho, \varphi)$  to
polar $ (\lambda, \omega)$ through the transformation:
$\rho=\epsilon (1+\lambda \sin \omega), \
\varphi=\epsilon(1+\lambda \cos \omega)$. The first set of
coordinates has the  range $(\rho>\epsilon , \varphi> \epsilon)$
and the second one\footnote{We ignore once again (since this is
sufficient for our purposes) some dependence of the region of
integration on $A$ and on higher powers of the cut-off $\eps$.}
$(\lambda> 0, \omega \in [ 0, {\pi \over 2}])$. Doing this
coordinate transformation in (\ref{region2}) and expanding in
powers of the cut-off $\epsilon$ we get:
\begin{equation}\label{region22}\epsilon^2 \int_0^{{\pi \over 2}} d
\omega\int_0^{\pi} d \theta \int_{0} d\lambda \ \epsilon^{4q \cdot
k_2 + 4 q \cdot k_1 + 4 k_1 \cdot k_2} (1+\lambda\sin\omega)^{4q
\cdot k_1-1} (1+ \lambda\cos \omega)^{4 k_1 \cdot k_2+1}
\end{equation}
where we took the zero momentum limit in the last term of
(\ref{region2}) after the change of variables. Using momentum
conservation we can show that the exponent of $\epsilon$ is equal
to two. The integral will be regular and cut-off independent in
the limit $\epsilon \to 0$. We can therefore take safely $A=0$
since all the singular terms come from
(\ref{region1}).\footnote{As we pointed out above, had we kept all
$A$ dependence in various formulas leading to (\ref{region22}) we
should have found a contribution cancelling the $A$-dependent
terms in the full integral.} Adding the contribution from the
$\varphi \in [{\pi \over 2}, \pi]$ region of integration we arrive
to the expression from (\ref{example}).

%%%%%%%%%%%%%%%%%%%%%%%%%%%%%%%%%%%%%%%%%%%%%%%%%%%%%%%5
%Bibliography


\begin{thebibliography}{99}
\bibitem{Fradkin}
E.~S.~Fradkin and A.~A.~Tseytlin, ``Nonlinear Electrodynamics From
Quantized Strings,'' Phys.\ Lett.\ B {\bf 163}, 123 (1985).
%%CITATION = PHLTA,B163,123;%%

\bi{Call}
A.~Abouelsaood, C.~G.~Callan, C.~R.~Nappi and S.~A.~Yost,``Open Strings In Background Gauge Fields,''
Nucl.\ Phys.\ B {\bf 280}, 599 (1987).
%%CITATION = NUPHA,B280,599;%%

\bi{poll}
J.~Dai, R.~G.~Leigh and J.~Polchinski,
``New Connections Between String Theories,''
Mod.\ Phys.\ Lett.\ A {\bf 4}, 2073 (1989).
%%CITATION = MPLAE,A4,2073;%%


\bi{POLL}
J.~Polchinski,
``Dirichlet-Branes and Ramond-Ramond Charges,''
Phys.\ Rev.\ Lett.\  {\bf 75}, 4724 (1995)
[arXiv:hep-th/9510017].
%%CITATION = HEP-TH 9510017;%%


\bibitem{LeighA}
R.~G.~Leigh,
``Dirac-Born-Infeld Action From Dirichlet Sigma Model,''
Mod.\ Phys.\ Lett.\ A {\bf 4}, 2767 (1989).
%%CITATION = MPLAE,A4,2767;%%

\bi{andr}
O.~D.~Andreev and A.~A.~Tseytlin,
``Partition Function Representation For The Open Superstring Effective Action: Cancellation Of Mobius Infinities And Derivative Corrections To
Born-Infeld Lagrangian,''
Nucl.\ Phys.\ B {\bf 311}, 205 (1988).
%%CITATION = NUPHA,B311,205;%%


\bibitem{malda}
S.~S.~Gubser, A.~Hashimoto, I.~R.~Klebanov and J.~M.~Maldacena,
``Gravitational lensing by $p$-branes,'' Nucl.\ Phys.\ B {\bf 472}
(1996) 231 [arXiv:hep-th/9601057].
%%CITATION = HEP-TH 9601057;%%

\bibitem{hashimoto} A.~Hashimoto and I.~R.~Klebanov,
``Scattering of strings from D-branes,'' Nucl.\ Phys.\ Proc.\
Suppl.\  {\bf 55B}, 118 (1997) [arXiv:hep-th/9611214].
%%CITATION = HEP-TH 9611214;%%



\bibitem{GarScatter} M.~R.~Garousi and R.~C.~Myers,
``Superstring Scattering from D-Branes,'' Nucl.\ Phys.\ B {\bf
475}, 193 (1996) [arXiv:hep-th/9603194].
%%CITATION = HEP-TH 9603194;%%

\bibitem{GarBI} M.~R.~Garousi and R.~C.~Myers,
``World-volume interactions on D-branes,'' Nucl.\ Phys.\ B {\bf
542}, 73 (1999) [arXiv:hep-th/9809100].
%%CITATION = HEP-TH 9809100;%%


%Derivative corrections
\bibitem{BachasCurvature}
C.~P.~Bachas, P.~Bain and M.~B.~Green, ``Curvature terms in
D-brane actions and their M-theory origin,'' JHEP {\bf 9905}, 011
(1999) [arXiv:hep-th/9903210].
%%CITATION = HEP-TH 9903210;%%

\bibitem{Wyllard}
N.~Wyllard, ``Derivative corrections to D-brane actions with
constant background  fields,'' Nucl.\ Phys.\ B {\bf 598}, 247
(2001) [arXiv:hep-th/0008125].
%%CITATION = HEP-TH 0008125;%%
 ``Derivative corrections to the D-brane Born-Infeld
action: Non-geodesic  embeddings and the Seiberg-Witten map,''
JHEP {\bf 0108} (2001) 027 [arXiv:hep-th/0107185].
%%CITATION = HEP-TH 0107185;%%


\bibitem{foto}A.~Fotopoulos,
``On $\a'^2$ corrections to the D-brane action for
non-geodesic  world-volume embeddings,'' JHEP {\bf 0109} (2001)
005 [arXiv:hep-th/0104146].
%%CITATION = HEP-TH 0104146;%%

\bi{TT}
A.~A.~Tseytlin,
``Heterotic - type I superstring duality and low-energy effective actions,''
Nucl.\ Phys.\ B {\bf 467}, 383 (1996)
[arXiv:hep-th/9512081].
%%CITATION = HEP-TH 9512081;%%


\bibitem{Corley}
S.~Corley, D.~A.~Lowe and S.~Ramgoolam, ``Einstein-Hilbert action
on the brane for the bulk graviton,'' JHEP {\bf 0107} (2001) 030
[arXiv:hep-th/0106067].
%%CITATION = HEP-TH 0106067;%%

\bibitem{ardalan}
F.~Ardalan, H.~Arfaei, M.~R.~Garousi and A.~Ghodsi, ``Gravity on
noncommutative D-branes,'' arXiv:hep-th/0204117.
%%CITATION = HEP-TH 0204117;%%



\bi{GKT}
S.~S.~Gubser, I.~R.~Klebanov and A.~A.~Tseytlin,
``String theory and classical absorption by three-branes,''
Nucl.\ Phys.\ B {\bf 499}, 217 (1997)
[arXiv:hep-th/9703040].
%%CITATION = HEP-TH 9703040;%%

\bi{GK}
S.~S.~Gubser and I.~R.~Klebanov,
``Absorption by branes and Schwinger terms in the world volume theory,''
Phys.\ Lett.\ B {\bf 413}, 41 (1997)
[arXiv:hep-th/9708005].
%%CITATION = HEP-TH 9708005;%%


\bibitem{witten98}
E.~Witten, ``Anti-de Sitter space, thermal phase transition, and
confinement in  gauge theories,'' Adv.\ Theor.\ Math.\ Phys.\
{\bf 2} (1998) 505 [arXiv:hep-th/9803131].
%%CITATION = HEP-TH 9803131;%%


\bibitem{tseytlin98} H.~Liu and A.~A.~Tseytlin,
``D = 4 super Yang-Mills, D = 5 gauged supergravity, and D = 4
conformal  supergravity,'' Nucl.\ Phys.\ B {\bf 533} (1998) 88
[arXiv:hep-th/9804083].
%%CITATION = HEP-TH 9804083;%%

\bibitem{GKP}
S.~S.~Gubser, I.~R.~Klebanov and A.~M.~Polyakov,
``Gauge theory correlators from non-critical string theory,''
Phys.\ Lett.\ B {\bf 428}, 105 (1998)
[arXiv:hep-th/9802109].
%%CITATION = HEP-TH 9802109;%%

\bibitem{Wi}
E.~Witten,
``Anti-de Sitter space and holography,''
Adv.\ Theor.\ Math.\ Phys.\  {\bf 2}, 253 (1998)
[arXiv:hep-th/9802150].
%%CITATION = HEP-TH 9802150;%%


\bibitem{sei}N.~Seiberg and E.~Witten,
``The D1/D5 system and singular CFT,'' JHEP {\bf 9904} (1999) 017
[arXiv:hep-th/9903224].
%%CITATION = HEP-TH 9903224;%%


\bibitem{kiritsis01} E.~Kiritsis, N.~Tetradis and T.~N.~Tomaras,
``Thick branes and 4D gravity,''
JHEP {\bf 0108} (2001) 012 [arXiv:hep-th/0106050].
%%CITATION = HEP-TH 0106050;%%
``Induced gravity on RS branes,''
JHEP {\bf 0203} (2002) 019 [arXiv:hep-th/0202037].
%%CITATION = HEP-TH 0202037;%%

\bibitem{Antoniadis02}
I.~Antoniadis, R.~Minasian and P.~Vanhove, ``Non-compact
Calabi-Yau manifolds and localized gravity,''
arXiv:hep-th/0209030.
%%CITATION = HEP-TH 0209030;%%

\bibitem{RS} L.~Randall and R.~Sundrum,
``A large mass hierarchy from a small extra dimension,''
Phys.\ Rev.\ Lett.\  {\bf 83}, 3370 (1999) [arXiv:hep-ph/9905221];
%%CITATION = HEP-PH 9905221;%%
``An alternative to compactification,''
Phys.\ Rev.\ Lett.\  {\bf 83}, 4690 (1999) [arXiv:hep-th/9906064].
%%CITATION = HEP-TH 9906064;%%



\bibitem{garriga} J.~Garriga and T.~Tanaka,
``Gravity in the brane-world,''
Phys.\ Rev.\ Lett.\  {\bf 84} (2000) 2778 [arXiv:hep-th/9911055].
%%CITATION = HEP-TH 9911055;%%

\bibitem{chiba} T.~Chiba,
``Scalar-tensor gravity in two 3-brane system,''
Phys.\ Rev.\ D {\bf 62} (2000) 021502 [arXiv:gr-qc/0001029].
%%CITATION = GR-QC 0001029;%%

\bibitem{POL}
J.~Polchinski,
``String Theory. Vol. 1: An Introduction To The Bosonic String,''
{\it  Cambridge Univ. Press (1998) 402 p}.


\bibitem{klebanov}
I.~R.~Klebanov and L.~Thorlacius, ``The Size of p-Branes,'' Phys.\
Lett.\ B {\bf 371}, 51 (1996) [arXiv:hep-th/9510200].
%%CITATION = HEP-TH 9510200;%%




\bibitem{Burgess}
C.~P.~Burgess and T.~R.~Morris, ``Open Superstrings A La
Polyakov,'' Nucl.\ Phys.\ B {\bf 291}, 285 (1987).
%%CITATION = NUPHA,B291,285;%%


\bibitem{bern}Z.~Bern,
``String based perturbative methods for gauge theories,''
arXiv:hep-ph/9304249.
%%CITATION = HEP-PH 9304249;%%

\bibitem{GNW} M.~T.~Grisaru, P.~van Nieuwenhuizen and C.~C.~Wu,
``Gravitational Born Amplitudes And Kinematical Constraints,''
Phys.\ Rev.\ D {\bf 12} (1975) 397.
%%CITATION = PHRVA,D12,397;%%

\bibitem{Kawai:1985xq} H.~Kawai, D.~C.~Lewellen and S.~H.~Tye,
``A Relation Between Tree Amplitudes Of Closed And Open Strings,''
Nucl.\ Phys.\ B {\bf 269} (1986) 1.
%%CITATION = NUPHA,B269,1;%%

\bibitem{dotsenko} V. Dotsenko, ''Serie de Cours sur la
Theorie Conforme'',
%(In French),
 LPTHE.



\end{thebibliography}
\end{document}